\documentclass[aps,preprint,showpacs,preprintnumbers,amsmath,amssymb,floats,groupedaddress]{revtex4-1}
\usepackage{graphicx}
\usepackage[english]{babel}
\usepackage{amsmath}
\usepackage{amssymb}
\usepackage{color}

\newcommand{\be}{\begin{eqnarray}}
\newcommand{\ee}{\end{eqnarray}}

\newcommand{\avs}{$A$V$_3$Sb$_5$}

\newcommand{\cvs}{CsV$_3$Sb$_5$}

\begin{document}

\title{Extended superconducting fluctuation region and 6e and 4e flux-quantization \\ in a Kagome compound with a normal state of 3Q- order}

\author{Chandra M. Varma}
\affiliation{Physics Department, University of California, Berkeley, CA 94704, USA\\
Physics Department, University of California, Riverside, Ca. 92521}
\thanks{Emeritus}

\author{Ziqiang Wang}
\affiliation{Department of Physics, Boston College, Chestnut Hill, MA 02467, USA}

\date{\today}
\begin{abstract}
The
superconducting state with the usual 2e-flux quantization formed from a normal state with 3Q charge density or loop-current order  is a linear combination of 3 different paired states with an overall gauge invariant phase and two internal phases such that the phases in equilibrium are at $2\pi/3$ with respect to each other. In the fluctuation regime of  such a 3-component superconductor, internal phase  fluctuations are of the same class as for frustrated classical xy- spins on a triangular lattice. The fluctuation region is known therefore to be abnormally extended below the mean-field or the Kosterlitz-Thouless transition temperature.    A 6e-flux and a 4e-flux quantized states can  be constructed which are also eigenstates of the BCS Hamiltonian and stationary points of the Ginzburg-Landau free-energy with a  transition temperature above that of the renormalized 2e-flux quantized state. Such states have no internal phases and so no frustrating internal phase fluctuations. These state however cannot acquire long-range order because their free-energy is higher than the co-existing fluctuating state of 2e flux-quantization.  6e as well as 4e- flux- quantized Little-Parks oscillations however occur in which the resistivity increases periodically with field above that of the 2e-fluctuating state in its extended  fluctuation regime, as are observed, followed at low temperatures to a condensation of the time-reversal odd 2e-quantized state
\end{abstract}

\maketitle

\section{Introduction}

The quasi-two dimensional Kagome lattice compounds $A$V$_3$Sb$_5$ \cite{stephen-prm, stephen-prl}, where $A=$ K, Rb, Cs are various alkalis, has attracted much attention recently for its normal state \cite{Hasan-nm21, Zeljkovic-nat21, XHChen-prx21, Li-prx21, Hasan-prb21, haihuwen-twofold, Zeljkovic-np22, Uykur-prb21, Ratcliff-prm21, Xie-prb22, Wu-prb22, Liu-nc22, sato, mingshi, comin, xinjiangzhou, musr-1, musr-2, musr-3, Wilson-prx21, Fu-prl21, Shrestha-prb22, kerr, kerr-liangwu, kerr-yonezawa, kerr-kapitulnik, Moll, pottsnematic-XHChen, Hongli-Unidirectional, MingShi-stacking, Comin-stacking, Kato-CommunMat22, ilija-pocket}
and for its superconductivity
\cite{HJGao-nat21,DLFeng-prl21, Ortiz-prm21, Yin-cpl21,SYLi-arXiv21, HQYuan-21, pressure1, pressure2, sn-doping, ti-doping, JiangWang-arXiv22, Kapitulnik-arXiv23} and the fluctuation regime above its
superconductivity.  It also has been the subject of many theoretical investigations \cite{binghai-prl, dmft, chiralflux, balents, Lin-prb21, Denner-prl21, Feng-prb21, miaohutheory, fernandes, ZhouWang, Fernandes-loopcurrent, Lin-Nandkishore, tv1v2, nematic-cdw}. The normal state has a transition to a 3Q structure at about $100~ K$ with possibly other transitions at lower temperature. The superconducting transition $T_c$ is between about 1 K to about 2.0 K depending on the samples. Experiments as well as theories \cite{ZhouWang, balents, Lin-prb21, Fernandes-loopcurrent, Lin-Nandkishore, tv1v2, nematic-cdw} have raised the possibility that the transition at $100~ K$ breaks time-reversal and chirality but preserves inversion and occurs to a state with loop-current order \cite{CMV1997}. This is however not a completely settled matter.  However as we will show, whether or not time-reversal is broken in the normal state is not crucial to the problem considered here. We are concerned here with the recent flux quantization experiments \cite{JiangWang-arXiv22} near and {\it above} the superconducting transition in the compound  \cvs  ~which are very surprising. Little-Parks type experiments in a ring geometry attached to leads have been performed. Starting at about 4 K, the resistivity begins to drop  as temperature is decreased much faster than above about 4 K. In this region, which is an abnormally extended region of superconducting fluctuations, flux quantization begins to be cleanly observed  but the flux quantum corresponds to charge $6e$. As temperature is decreased but still above $T_c$, the quantization gets a little muddier with $6e$, $4e$ as well as the usual $2e$ discernible. Very close to $T_c$ and below only $2e$ quantization is observed.

These are extraordinary results. There are four aspects to them to be understood: First, why is the fluctuation region so extended? Second, why  is flux quantization observed so far above $T_c$ and why does it correspond to 6e-flux quantization? Third, despite the second, why is the ultimate superconducting state of the usual 2e flux quantization? Fourth, why do the  fluctuations in resistivity occur without the resistivity going to zero, as in the usual Little-Parks oscillations?

\begin{figure}[ht]
 \begin{center}
 \includegraphics[width= 1.0\columnwidth]{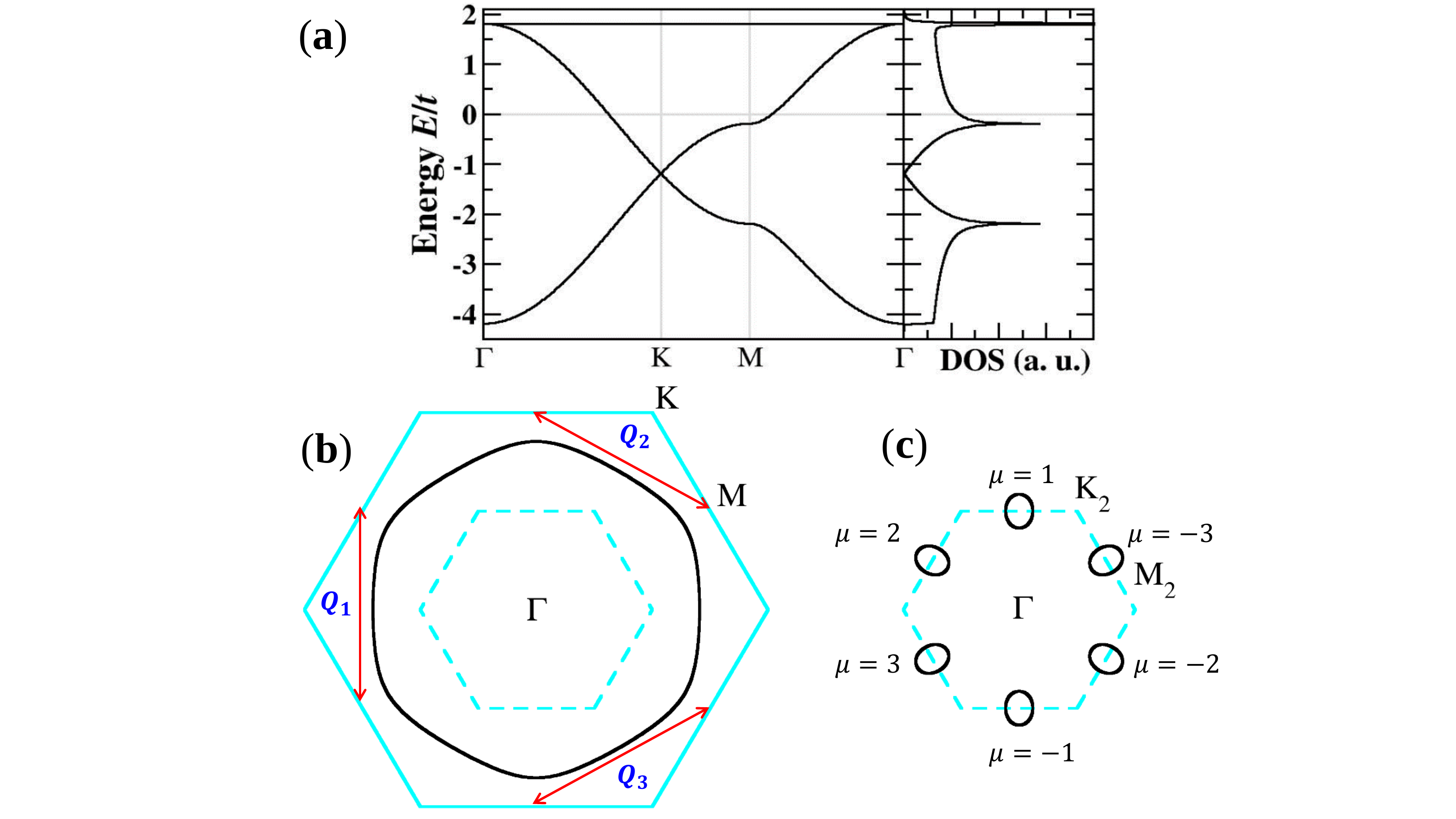}
 \end{center}
\caption{Simple model illustrating the 6 elliptical Fermi-surface pockets below the charge density or loop-current order transition proposed for \avs. (a) The band structure of the simplest one-orbital tight-binding model with nearest neighbor hopping on the kagome lattice. The Fermi level is placed just above the van Hove singularity at the M points, corresponding to the Fermi surface marked by the black line in (b). The van Hove points are connected by the three wave vectors ${\bf Q}_i$, $i=1,2,3$ (red arrowed lines) along the three hexagonal directions. ${\bf Q}_i={1\over2}{\bf G}_i$ and ${\bf G}_i$ are the reciprocal lattice vectors, which is responsible for the propensity toward $2\times2$ bond ordered 3Q CDW.
In (b), the larger hexagon (solid cyan line) is the $1\times1$  Brillouin zone, while the smaller hexagon (dashed cyan line) marks the reduced Brillouin zone in the enlarged $2\times2$ unit cell. (c) The reconstructed Fermi surface pockets plotted in the reduced Brillouin zone due to the $2\times2$ CDW order. The superconducting states are constructed from the three possible Cooper pair states of the fermions in a given ellipse and its inversion related ellipse ($\mu=\pm1,\pm2,\pm3$), giving rise to a three-component superconductor.}
 \label{Fig:FS}
\end{figure}

The organization of this  paper and the principal results are as follows. In Sec. II, we summarize what is known in experiments and in theory about the normal state of $A$V$_3$Sb$_5$ below the transition at about 100 K. We take the simplest model which in the normal state has 6 small elliptical Fermi-surface pockets, as schematically illustrated in Fig.~\ref{Fig:FS}. In Sec. III, we consider the superconductive state with 2e flux quantization with a BCS reduced Hamiltonian and emphasize that it is a linear combination with complex coefficients of the three zero center of mass momentum states made of the three pair states of fermions from inversion related elliptical pockets.  We also consider in this section possible 6e flux quantized states with the BCS reduced Hamiltonian. The next section considers the fluctuations of the 2e states using the Ginzburg-Landau type free-energy and shows that the fluctuations in this state are quite unusual because the internal phase fluctuations between the three cooper-pair states are mapped to a model of frustrated classical spins on a triangular lattice. Using the work done on the latter  long ago \cite{Shiba, Babaev2011, Stanev2012, Chubukov2013, Sudbo2013, Sudbo2014, Yanagisawa}, we argue that this introduces a very large temperature region of chirality  and phase fluctuations  in which the resistivity decreases rapidly with temperature.

\begin{figure}
 \begin{center}
\includegraphics[width= 1.0\columnwidth]{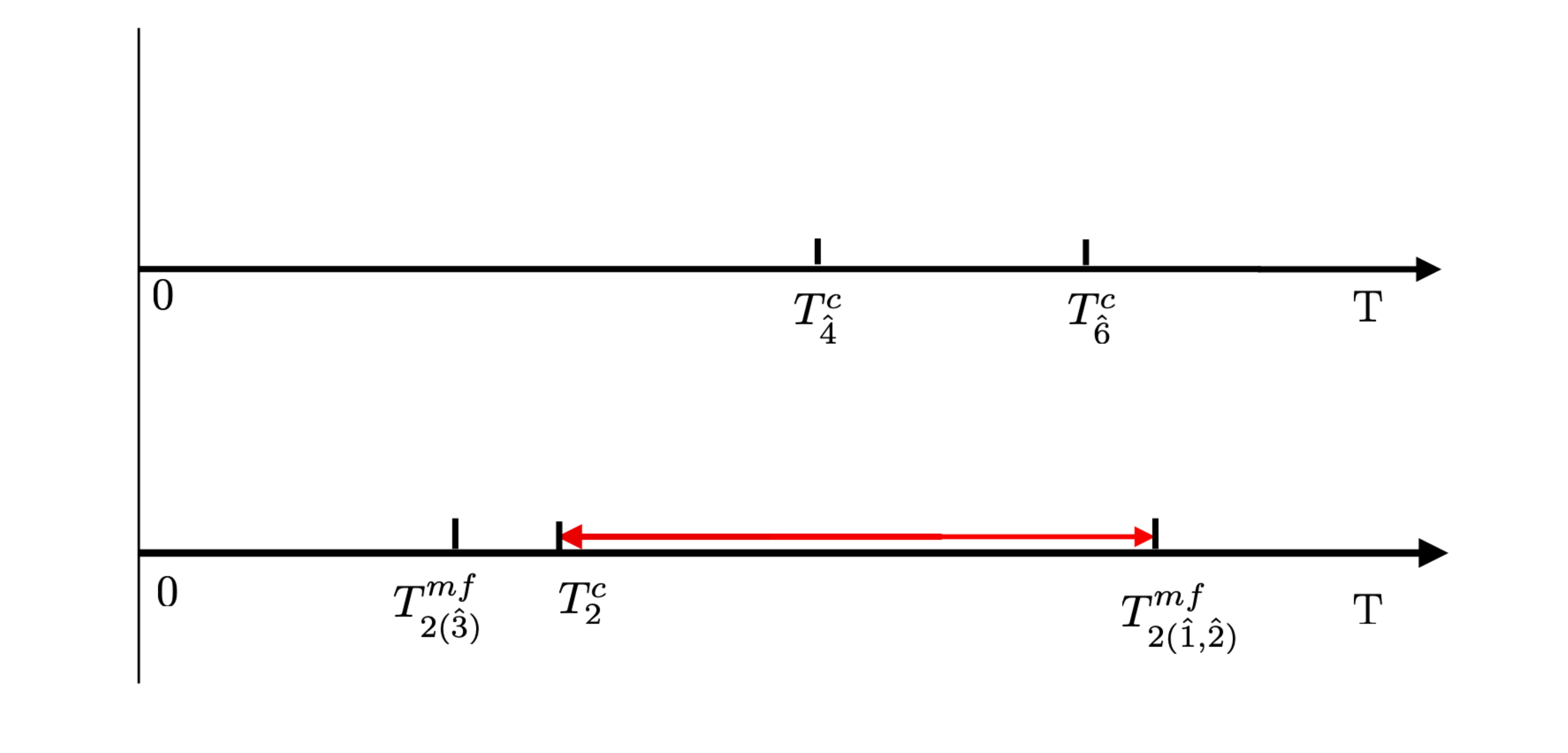}
\end{center}
\caption{The figure sketches the various characteristic temperatures which are of importance in the results and discussions in this paper.
 $T_{2e}^{c}$ is the actual transition temperature to a 2e-flux quantized time-reversal breaking superconducting state.
 $T^{mf}_{2e, (\hat{1},\hat{2})}$
 are the mean-field transition temperature of the two degenerate time-reversal breaking $2\pi/3$ phase difference states with 2e flux quantization, while $T^{mf}_{2e, \hat{3}}$ is the mean-field transition temperature of the real state with the same quantization.
 The region between $T^{mf}_{2e,(\hat{1},{\hat{2}})}$ and $T_{2e}^{c}$, marked in red, is the extended region of chirality and phase fluctuations. From Monte-carlo calculations of \cite{Shiba}, $T_{2e}^c \approx 0.502 J$, where $J$ is the Josephson coupling energy of the pairs of phases of the three components from pockets $(\mu, -\mu), \mu =1,2,3$ of the superconducting state shown in Fig. (\ref{Fig:FS}). $T^c_{\hat{6}e}$ and $T^c_{\hat{4}e}$ are the transition temperatures to the 6e and 4e flux quantized states described in the text. The parameters through which the actual values of these temperatures are determined are not known. Arguments are given in the text for their approximate relative placements. The 4e and the 6e states never lead to zero resistive states because the fluctuating region of the 2e flux quantized states have a lower free-energy and co-exist with them. However, resistivity fluctuations with 6e and 4e flux quantization in a magnetic field through the ring occur on top of the resistivity of the 2e fluctuating states.}
 \label{Fig:Temperatures}
\end{figure}

 The abnormally large  fluctuations reduce the condensation temperature to such a state drastically in relation to the mean-field BCS transition temperature. A  6e quantized state and a 4e quantized state can be constructed which are also eigenstates of the BCS reduced Hamiltonian and which are unfrustrated.  However the free-energy of such a state is actually higher than the fluctuating 2e-states, which are mutually orthogonal to the 6e and 4e states. Therefore a product state of the three must form. Such a state is shown to have Little-Parks oscillation with 6e quantization  and 4e quantization on top of the background of  the sharply changing resistive state of the fluctuating 2e states. Long-range order  occurs in the 2e state at a lower temperature below which the other states disappear. For convenient reference to the rest of the paper, the various important temperatures are sketched in (\ref{Fig:Temperatures}).
 
\section{3Q charge or Loop-current ordered state in $ AV_3Sb_5$}

We consider the minimal single-orbital model with nearest neighbor hopping. The band structure is shown in Fig. 1a in the Brillouin zone plotted in Fig. 1b. The van Hove singularities at the M points are connected by
the three vectors ${\bf Q}_i={1\over2}{\bf G}_i$, where ${\bf G}_i$, $i=1,2,3$ are the reciprocal lattice vectors as shown in Fig. 1b. As a result,
this band-structure is well nested at van Hove filling. On this band-structure the projected one-particle states have zero on-site interactions and finite values for nearest and next-nearest neighbor interactions \cite{jxli, thomale-prb, qhwang, thomale-prl, tv1v2}. This is especially propitious for 3Q ordered states, with or without time-reversal breaking, which have a $2\times2$ enlarged unit cell and a folded Brillouin zone with the
area reduced by a factor of four as shown in Fig. 1b.

The Fermi level of \avs ~ is away but close to the van Hove singularities. The case where the Fermi level is just above the van Hove filling is shown in Fig. 1a with the corresponding Fermi surface plotted in Fig. 1b.
The favored $2\times2$  loop current or real 3Q-CDW states give rise to 6 reconstructed small elliptical pockets centered at the M-points in the reduced zone \cite{ZhouWang, tv1v2} as shown in Fig. 1c. Let $c_{\bf k}$ denote the annihilation operators for the band near the chemical potential in the high temperature phase. The quasiparticle annihilation operator $a_{\bf k, \mu}$ on the three-elliptical pockets and their time-reversed and inversion related states are given by
\be
\label{ann}
a_{{\bf k, \mu}} = \sum_{ i = 1,2,3} |u_{k , \mu, {\bf Q}_i}| e^{i \theta_{{\bf k}, \mu, {\bf Q}_i}} c_{\bf k + {\bf Q}_i}.
\ee
$\mu = \pm(1,2,3)$ are the indices for the three-ellipses (plus sign) and their inversion related partners (minus sign). We will ignore the spin-indices throughout; the pairing will be understood to be in the spin-singlet channel in whichever even angular momentum is favored. Our results do not depend on the details of the pairing symmetry. The difference in the phase factors $\theta$ at different ${\bf Q}$ is responsible for the time-reversal breaking and chirality. We can safely ignore the ${\bf k}$-dependence of the phase factor because ${\bf Q}_i$ are much larger than the size of the elliptical pockets. Also preserving inversion is equivalent to
\be
\label{inv}
\theta_{\mu_i, {\bf Q}_i} = \theta_{-\mu_i, -{\bf Q}_i} \equiv \theta_i.
\ee
If the loop-current order is chiral, the three $\theta_{\mu_i, Q_i}$ are unequal.

 If time-reversal symmetry is indeed broken in the normal state, it is important to consider the energy of spatial variation of the two internal fluctuating phases $\theta_{12} \equiv (\theta_1 - \theta_2)$ and of  $\theta_{13} \equiv (\theta_1 - \theta_3)$. This energy is determined by the effective interactions of fluxes in adjacent cells.  The energy of similar modes has been calculated in other contexts \cite{He_V2011, He-V2014}). The 
 excitation energy for such modes for long wave-length fluctuations may be written as
 \be
 \label{fluc}
 \Omega(q) = \Omega_0 + J q^2,
  \ee
 where  both $\Omega_0$ and $J$ are on the scale of the transition temperature of the loop-current order. This is about two orders of magnitude larger than the superconducting transition temperature so that the equilibrium values of $\theta_{12}$ and $\theta_{13}$ may be considered fixed to the normal state value in the vicinity of the superconducting state. As worked out in detail below, these phases are inherited by the internal phases between pairs made from different Fermi-surface pockets in Fig. 1c. The large energy for variation as in Eq. (\ref{fluc}) make such phase variations in the superconducting state very stiff. For this reason, the dynamics of the loop-current order is not an essential aspect in the considerations below

\section{Superconductive states}
\subsection{2e- quantized flux superconducting states}

The superconducting states can  be constructed from single-particle states given by the  operators in (\ref{ann}), i.e. by forming Cooper pairs on inversion related elliptical pockets $(\mu_i, -{\mu}_i), i =1,2,3$. We do not consider paired states formed on pockets labeled by different $\mu$, which carry finite center of mass momentum
$\Delta {\bf Q}_{ij}={\bf Q}_i-{\bf Q}_j$ and cost the large additional kinetic energy $\sim (\Delta {\bf Q}_{ij})^2$. Although a weak spatial modulation of the superconducting state at wave vector $\Delta {\bf Q}_{ij}$ has been detected by STM in \cvs~\cite{HJGao-nat21}, we consider it a secondary effect brought about by coupling of the amplitude of pairing between ellipses centered at different {\bf
Q} and irrelevant to  our mechanism for the 6e and 4e flux quantization originates entirely by considering only the possible uniform superconducting states.

Consider the BCS reduced Hamiltonian
\be
\label{Hred}
H =& H_0 + H',
\ee
where $H_0$  describes the zero center of mass momentum pairing between states in $\mu$ and in $-{\mu}$,
\be
 \label{H0}
H_0 &=& \sum_{\mu=1,2,3} H({\mu, -\mu}), \\ \nonumber
H({\mu, -\mu}) &=& \sum_{{\bf k}} \big(\epsilon_{{\bf k}, \mu}
+ \epsilon_{-{\bf k},- \mu} - 2\zeta_0) a^+_{{\bf k}, {\mu}}a_{{\bf k}, {\mu}} + U  \sum_{{\bf k, k}'} a^+_{{\bf -k},-{\mu}} a^+_{{\bf k}, {\mu}}a_{{\bf k'}, {\mu}}a_{{\bf -k'}, -{\mu}}.
\ee
and $\zeta_0$ is the chemical potential. The Cooper-pair  annihilation operator is then
\be
\label{cooann}
b_{{\bf k}, \mu} \equiv a_{{\bf k} \mu} a_{-{\bf k} -\mu} &=&  e^{2 i \theta_{\mu}}
\sum_{i=1,2,3} |u_{{\bf k} , \mu, {\bf Q}_i}||u_{-{\bf k} ,-\mu, {\bf -Q}_i}|c_{{\bf k}+{\bf Q}_i}c_{-{\bf k}-{\bf Q}_i}.
\ee
Here we have used the relations (\ref{inv}) to define $2 \theta_{\mu}$.
$H'$ describes the scattering between the Cooper pairs formed on $\pm\mu$ and $\pm\mu^\prime$ pockets,
\be
\label{interband}
H'(\mu, -\mu; \mu', -\mu') =  \sum_{\mu\ne \mu'}{\sum_{{\bf k},{\bf k'}}}V_{\mu, \mu'}  e^{2 i (\theta_{\mu} - \theta_{\mu'})} b^+_{{\bf k}, \mu} b_{{\bf k'}, \mu'} + H.C.
\ee
The phase factor in (\ref{interband}) is intrinsic and any gauge transformation must preserve it. For the hexagonal symmetry of the states on the Kagome lattice, we consider
$V_{\mu \ne \mu'} \equiv V$.
The pairing with flux quantization 2e will in general be in a linear combination of the $b_{{\bf k}, \mu}$ operators
\be
\label{linb}
\hat{b}_{{\bf k}, \hat{\mu}} = \sum_{\mu =1,2,3} A_{\hat{\mu}, \mu} b_{{\bf k}, \mu}, ~\hat{\mu} = 1,2,3.
\ee
with orthonormal $A_{\hat{\mu}, \mu}$,
\be
\label{A}
\sum_{\mu}  A^*_{\hat{\mu}, \mu}A_{\mu, \hat{\mu}'} = \delta_{\hat{\mu},\hat{\mu}'}.
\ee
$A_{\hat{\mu}, \mu}$ includes the effects of the intrinsic phases $2\theta_{\mu}$ as well as the Josephson phases.

\subsection{Microscopics}

The energy of the three $\hat{\mu}$ states, each of them with 2e flux-quantization, will in general be different. Let us denote the three BCS states constructed from them as $\Psi_{\hat{\mu}}$,
\be
\label{BCS2}
\Psi_{\hat{\mu}} = \Pi_{{\bf k}} \big(\cos \theta_{\bf k, \hat{\mu}} + \sin \theta_{\bf k, \hat{\mu}} \hat{b}^+_{{\bf k}, \hat{\mu}}\big) |FS \rangle.
\ee
Recall that $\hat{b}^+_{{\bf k}, \hat{\mu}}$ is a sum of three terms as given in Eq. (\ref{linb}) with three different phases. It is important to note that after taking out an overall phase factor in $A_{\hat{\mu}, \mu}$, which gives the usual phase factor in $\Psi_{\hat{\mu}}$ responsible for the 2e-flux quantization, there remain two internal relative phase factors in $\hat{b}^+_{{\bf k}, \hat{\mu}}$. They come from the fixed relative phases of the normal state basis wave-functions between states at different ${\bf Q}_i$. They cannot be removed. If we represent the wave-function by spin-operators as in the basis used by Anderson, these would specify the two relative orientations of the spin-operator representations of the pairing operator.  As has been realized \cite{Shiba, Babaev2011, Stanev2012, Chubukov2013, Sudbo2013, Sudbo2014, Yanagisawa}, the relative phase factors in 2e-quantized flux states act as (resolvable) frustrations. Let us denote the three possible ground state energies by $E^0_{\hat{\mu}}$.

The magnitude of the gaps as a function of temperature and the relative phase of the gap function as well as $T_c$, i.e. $\theta_{k, \hat{\mu}}$ as well as $A_{\hat{\mu}, \mu}$ are given by a slight generalization of the BCS equation to a $3 \times 3$ matrix,
\be
\label{gapeqn}
\Psi_{\hat{\mu}} &=& \sum_{\hat{\mu}'} \hat{V}_{\hat{\mu}, \hat{\mu}'} N_{\mu'}\zeta_{\mu'} \Psi_{\hat{\mu}'} ,\\
\zeta_{\mu} &=& \int_0^{\omega_c} d\epsilon_{\mu} \tanh \frac{E_{\hat{\mu}}}{2k_BT}.
\ee
%
Here, $N(\mu)$ are the density of states of the $\mu$-th Fermi-surface, which are all equal in our problem. The diagonals give the usual BCS equation for each of the three components in the un-coupled limit where $\Psi_{\hat{\mu}} = \Psi_{\mu}$. As will be discussed in the next section, in the symmetry of the present problem, there are
two degenerate time-reversal odd superconducting states and a non-degenerate time-reversal even superconducting state. Which of the two has a higher mean-field transition temperatures depends on the sign of $V$. The sum of the three BCS transition temperatures is independent of $V$.

\section{Ginzburg-Landau Analysis for the superconducting fluctuations}

For discussing  flux quantization and especially Little-Parks effect in the vicinity of the phase transition and in the fluctuation regime above it, it is more convenient to adapt the Ginzburg-Landau (GL)  formalism for the problem. We adopt the convention that a basis state in the GL free-energy denoted by $\Phi_{\mu}$ transforms to the state  $\Psi_{\mu}$ in an ordered state which is described by BCS theory.  In a two-dimensional situation above the transition, the phase fluctuations determine the correlation functions. In that regime, i.e. below the mean-field transition temperature the amplitudes may be given a fixed value 
$|\Phi_{\mu}|$ since their fluctuations are unimportant. The leading GL free-energy for the phase fluctuations $\phi_{\mu}({\bf r})$  for the case of the 6 elliptical pockets at $\pi/3$ with respect to each other in momentum space and pairing in $\mu$ states with $(\mu,\mu')$ coupling as discussed above,
\be
\label{LG}
F &=& \sum_{\mu} \alpha_{\mu} |\Phi_{\mu}|^2 
+ 
 \sum_{\mu \ne \mu'} \alpha_{\mu,\mu'} \big(\Phi^*_{\mu}({\bf r})\Phi_{\mu'}({\bf r}) + H.C.\big)  + \gamma^2 |(-i \nabla -\frac{e}{c} {\bf A}) \Phi_{\mu}({\bf r})|^2, \\
\Phi_{\mu} &=& |\Phi_{\mu}|e^{i \phi_{\mu} ({\bf r})},\\
\alpha_\mu &=& \alpha \equiv  \alpha_0 \frac{T-T_0^{mf}}{T_0^{mf}}, ~
\alpha_{\mu,\mu'}  \equiv  \alpha' e^{2i(\theta_{\mu} -\theta_{\mu'})}.
\ee
The first term in (\ref{LG}) is the diagonal free-energy for the three $\mu$ and the second term is their mutual Josephson coupling energy.  The equality of the three $\alpha_{\mu}$ is due to the symmetry in the problem and $\alpha'$ being the magnitude of the Josephson coupling is real.

The multi-band free-energy  (\ref{LG}) has been analyzed extensively by Tanaka and Yanagisawa for its properties \cite{tanaka-1}, including the collective modes in the superconducting states \cite{Yanagisawa}. 
The $\Phi$'s and their spatial variations are given by minimizing Eq. (\ref{LG}). The result for zero external potential ${\bf A}$ is
\be
\label{LGEq}
\left(\begin{array}{ccc} \alpha  & \alpha_{12} & \alpha_{13} \\  \alpha_{12}^* & \alpha  & \alpha_{23} \\ \alpha_{13}^* & \alpha_{23}^* & \alpha  \end{array}\right)\left(\begin{array}{c}\Phi_{1} \\\Phi_{2} \\\Phi_{3}\end{array}\right) = \big( \gamma^2 \nabla^2 \hat{I}\big) \left(\begin{array}{c} \Phi_{1} \\  \Phi_{2} \\  \Phi_{3}\end{array}\right) .
\ee
The mean-field transition temperatures are given by setting the right side to $0$ and equating the eigenvalues of the left side to $0$.  There is a doubly degenerate eigenvalue: $E_{1,2} = \alpha - \alpha'$ and one nondegenerate eigenvalue $E_3 = \alpha +2 \alpha'$.  The eigenvalues do not depend on whether $\alpha_{ij}$'s are real as for the case that the normal state is a real charge density wave state, or whether the normal state has loop-current order so that the $\alpha$'s are complex.  The mean-field transition temperatures are changed from $T^0_{mf}$ when $\alpha' =0$ to
\be
T_{2e, \hat{1}, \hat{2}}^{mf} &=& T^{mf}_0 (1+ \frac{\alpha'}{\alpha_0}), \quad {\rm doubly~degenerate}, \\
T_{2e,\hat{3}}^{mf} &=& T^{mf}_0 (1- 2\frac{\alpha'}{\alpha_0}), \quad {\rm nondegenerate}.
\ee
Which state wins for the mean-field transition temperature depends on the sign of Josephson coupling. These temperatures are sketched in Fig. (\ref{Fig:Temperatures}) for the case that the time-reversal breaking states have a higher transition temperature. The flux quantization of all three is in units of 2e.
The eigenstates in the ordered state for the doubly degenerate solution, denoted by $\Phi_{\hat {1}}$ and $\Phi_{\hat{2}}$ are time-reversal odd with $\Phi^*_{\hat {1}} =  \Phi_{\hat {2}}$.
They are given in the basis $(\Phi_1, \Phi_2, \Phi_3)$ by
\be
 \Phi_{\hat{1}, \hat{2}}  = \frac{1}{\sqrt{3}} e^{i \phi} (1,e^{\pm 2 i\pi/3}, e^{\mp 2 i \pi/3}).
\ee
The non-degenerate state $\Phi_{\hat{3}}$ is real with the relative phases of the three superconducting components locked,
\be
 \Phi_{\hat{3}} = \frac{1}{\sqrt{3}} e^{i \phi} (1,1,1)
 \ee
We have taken out a gauge invariant phase which couples to the externally applied magnetic field. We will not have to deal with it till we discuss the flux quantization.

For the case of the loop-current ordered normal state, eigenvectors are
\be
\label{ev/clco}
E_3&=& (\alpha +2\alpha'), \quad \Phi = \frac{1}{\sqrt{3}} e^{i \phi} (e^{-2 i \theta_1},e^{-2 i \theta_2},e^{-2 i \theta_3}) \\
E_{1,2}& =& (\alpha-\alpha'), \quad \Phi  = \frac{1}{\sqrt{3}} e^{i \phi} (e^{-2 i \theta_1},e^{-2i \theta_2 \pm 2 i\pi/3}, e^{-2i \theta_3 \mp 2 i \pi/3}).
\ee
For the chiral case, $\theta_1 \ne \theta_2 \ne \theta_3$, all three states break time-reversal invariance in the superconducting state. For the non-chiral case,
$\theta_1=  \theta_2 = \theta_3$, we can take the $\theta$'s out and add to the gauge invariant phase. Actually the superconducting fluctuations for this state are similar to that for the real 3Q-CDW, given the stiffness of the loop-current fluctuations. We shall therefore henceforth only discuss
$ \Phi_{\hat{1}, \hat{2}}$ and $\Phi_{\hat{3}}$.

\subsection{Fluctuations of the 2e-states}

 It is appropriate to consider the fluctuations to the superconducting state in \cvs\ to be two-dimensional.  In that case, the important fluctuations are in the phase variable with the amplitude varying slowly with temperature so that we can take it to be fixed. There are three phases $\phi_i({\bf r})$, defined by
 \be
 \Phi_{\hat{i}}({\bf r})  = |\Phi_{\hat{i}}| e^{i \phi_i({\bf r})}.
 \ee
One can take out an over-all phase for the wave-functions, say the sum of the three, to which the external field couples and which we have denoted as $\phi$ above. That leaves two internal phase variables for the fluctuations. The equation for the fluctuations of the superconducting state $\Phi_{\hat{i}}$ in zero external field obtained from Eq. (\ref{LGEq}) is then
\be
\label{deq}
J \nabla^2 \phi_i  ({\bf r})=  \alpha'\sum_{j \ne i}\cos(\phi_i - \phi_j)({\bf r}),
\ee
where, in terms of parameters introduced above, $J = \gamma^2$ is the stiffness  and $\alpha'$ is the Josephson coupling. As has been  discussed \cite{Shiba, Babaev2011, Stanev2012, Chubukov2013, Sudbo2013, Sudbo2014, Yanagisawa}, this model maps to a model of classical spins on a triangular lattice. 
The multi-band free-energy  (\ref{LG}) has been analyzed extensively by Tanaka and Yanagisawa for its properties \cite{tanaka-1}, including the collective modes in the superconducting states \cite{Yanagisawa}. 

We borrow in the next section some results for the fluctuation regime above the superconducting transition temperature which are important for us, on a related model  studied by Miyashita and Shiba \cite{Shiba}. Miyashita and Shiba \cite{Shiba} do the calculations on a discrete triangular lattice, which may be identified with the three distinct sites on a triangular lattice associated with the pair of bands $(\mu, -\mu)$. In that case we may define the problem as a problem of xy model on a triangular lattice with an effective Hamiltonian
\be
\label{t-lat-xy}
H = J \sum_{i \ne j = 1,2,3} \cos\big(\phi_i({\bf r}) - \phi_j({\bf r})\big).
\ee
The model has been investigated in detail by Miyashita and Shiba \cite{Shiba} by Monte-Carlo methods.  In the continuum approximation, minimization with respect to the phases of (\ref{t-lat-xy}) leads to Eq. (\ref{deq}).
Let us clearly restate the sense in which the  equation (\ref{deq}) derived from the GL Hamiltonian maps to the model solved in Monte-Carlo calculations of the xy model on a triangular lattice \cite{Shiba}.
The latter is a discrete lattice model in which at each lattice site a vector of fixed length lies in an arbitrary direction in the disordered state well above the fluctuation regime. Just below the mean-field transition temperature where short range order develops, the vectors on a nearest neighbor triangle lie at $2\pi/3$ with respect to each other but the triad's direction in nearby triangles are disordered. The fluctuation regime consists of ordering of these triads as temperature decreases. Eq. (\ref{deq}) is a continuum equation in which a coarse graining of the triangular lattice has been performed; it is valid only in the  above fluctuation regime. At each point ${\bf r}$ in the continuum, three vectors $\phi_i({\bf r})$ exist which lie at  $2\pi/3$ with respect to each other. The nabla operator refers to variations on spatial scale much larger than the triangular lattice constant. The eventual ordering is the relative ordering of the triads at arbitrarily long length scale, which is the same in the discrete as well as the continuum model.

The model with ferromagnetic $\alpha'$ or $J$ is unfrustrated. We are concerned only with the more interesting  antiferromagnetic $\alpha'$ which introduces frustration.  At equilibrium, the three phases are at $2\pi/3$ with respect to each other. There are two ways to realize that, with opposite chiralities. So besides fluctuations characteristic of the XY model, there are also fluctuations of chirality which are of the Ising class \cite{Shiba, Babaev2011, Stanev2012, Chubukov2013, Sudbo2013, Sudbo2014, Yanagisawa}. An Ising model on a triangular lattice is frustrated and so there is a much larger region in temperature of fluctuations than in an XY model alone. In general there are two phase transitions, separated by a  temperature unmeasurable in the Monte-Carlo calculations.
Unlike the phase transition of the XY model which shows essentially no specific heat singularity, the specific heat for the model shows a logarithmic singularity characteristic of the
Ising model in two dimensions. Above the phase transition to a phase with $2\pi/3$ differences in the three $(\phi_i- \phi_j)$ with a chosen chirality, thermal entropy favors a temperature region in which one of the three phases fluctuates about $0$ so that the frustration is removed for the other two phases as they can be at $\pi$ with respect to each other \cite{Yanagisawa}.

For our purposes, it is important to deduce from the references given above, the extent of the fluctuation regime and the decrease of the temperature of the transition they cause from mean-field BCS transition or the transition temperature of the unfrustrated KT transition in which at every point ${\bf r}$ only one-vector lies and not a triad.  Miyashita and Shiba \cite{Shiba} estimate that the transition temperature in the frustrated model is about $0.502 J$, while the Kosterlitz-Thouless transition temperature of the unfrustrated model is about $0.95 J$ \cite{Miyashita}. In Fig. (\ref{Fig:Temperatures}), the former is denoted by $T_{2e}^c$.  The latter itself is always lower than the BCS transition temperature, the ratio of the two depends on details of the interactions and are typically about 1/2. So we expect a fluctuation regime in \cvs\ which is two to four times the actual transition temperature. The fluctuation region shown by the red-line in the same figure extends from $T_{2e,(\hat{1},\hat{2})}^{mf}$. The latter for an xy model on a triangular lattice is $1.5 J$ \cite{Shiba}. This is in qualitative accord with the fact that the resistivity begins to drop in thin film samples of \cvs\ at about $4 K$ and the transition temperature to the zero-resistance 2e-superconducting state occurs at about 2 K in the best samples. The transition temperature is lower in ring geometry samples with the lowest at about $1 K$ and the fluctuation resistivity starting in all samples at about $4 K$. No Little-Parks oscillation are possible with 2e flux quantization in the fluctuation regime because the amplitude of such oscillations is proportional to
$\xi(T)^2/R^2$ \cite{Tinkham}, where $\xi(T)$ is the superconducting correlation length and $R$ is the radius of the ring.

\noindent
\subsection{6e-and 4e-flux quantized states}

A state with charge-6e flux quantization
is simply the product of the three orthogonal 2e-flux states which we have considered above
\be
\label{6}
\Phi_{\hat{6}} = \Pi_{\hat{\mu} = 1,2,3} \Phi_{\hat{\mu}} = |\Phi_{\hat{1}}|^2 \Phi_{\hat{3}} = | \Phi_{\hat{2}}|^2 \Phi_{\hat{3}}.
\ee
The second two equalities follow from the fact that $\Phi^*_{\hat{1}} = \Phi_{\hat{2}}$.

It is also straight-forward to write a BCS state with 6e-flux in terms of the notation introduced in Eq. (\ref{cooann}).
\be
\Psi_{\hat{6}} = \Pi_{\hat{\mu}=1,2,3} \Pi_{{\bf k}}\big(\cos \beta_{{\bf k},\hat{\mu}} + \sin \beta_{\bf{k},{\hat{\mu}}} {b}^+_{{\bf k}, {\hat{\mu}}}\big) |FS>.
\ee
Its BCS transition temperature $T^{mf}_{{\hat{6}}}$ is a third of the sum of the transition temperatures of the three 2e-flux states and so given simply by $\alpha = 0$, i.e. below the  mean-field transition temperature $T^{mf}_{2e, \hat{2}, \hat{3}}$ of the time-reversal breaking states by a factor $(1- 2 \alpha'/\alpha)$. However, we have a large regime of parameters in which $T^{mf}_{\hat{6e}}$ is larger than the KT transition temperatures $T^c_{\hat{2}, \hat{3}}$ of the states $\Phi_{\hat{2}, \hat{3}}$.

From Eqs. (\ref{6}), we gather that $\Phi_{\hat{6}}$ is purely real, except for an overall multiplicative factor  $e^{i \phi_6({\bf r})}$ which couples to an external field. Since it has no internal phase fluctuations, its fluctuation regime is just that for an ordinary two-dimensional superconductor and not an extended fluctuation regime as for the chiral $\Phi_{\hat{1}, \hat{2}}$ states. $\Phi_{\hat{6}}$ is quantized by charge-6e flux quantum. But this state cannot exist by itself because its free-energy below $T^c_{\hat{6e}}$ is larger than that of the states obtained from $\Phi_{\hat{1}, \hat{2}}$ which, including their fluctuations, reduce the free-energy by $ - TS(T)$, where $S(T)$ is the entropy of the fluctuations.

Similarly a uniform 4e-flux state
can be obtained by a product of the two time-reversal breaking 2e states:
\be
\label{4}
\Phi_{\hat{4}} = \Pi_{\hat{\mu} = 1,2} \Phi_{\hat{\mu}} = |\Phi_{\hat{1}}|^2 = | \Phi_{\hat{2}}|^2.
\ee
This is an eigenstate of the BCS Hamiltonian and an extremum of the Ginzburg-Landau free-energy. Being a pure real state in its internal co-ordinates, it has only the usual Kosterlitz-Thouless fluctuations of the overall  phase variable. The BCS transition temperature of such  a state is the same as the BCS transition of the 2e state. The ratio of the KT transition temperatures for the 4e and 6e states, which is more relevant, is discussed in the next section.

It is important also to note that the state $\Phi_{\hat{6}}$ is not orthogonal to the state which is the product of another $6e$ quantized state $\Phi_{{6}} \equiv  \Phi_1 \Phi_2 \Phi_3$, i.e to the product of the states defined in (\ref{LG}) which are the basis states in Eq. (\ref{LGEq}). The latter ignore the off-diagonal couplings $\alpha_{\mu,\mu'}$ and have higher energy. Similarly the state $\Phi_{\hat{4}}$ is not orthogonal to unstable states  $\Phi_1 \Phi_2$. Therefore the states $\Phi_{\hat{6}}$ and $\Phi_{\hat{4}}$ are never stable and must be considered as decaying in time. 

\section{Little-Parks Oscillations with charge $6e$ flux quantization}

We now come to the experiments \cite{JiangWang-arXiv22} which motivated these investigations.
Let us denote the state in the extended phase fluctuation temperature region  by $\Phi_{2e-fl}(T)$.
$<\Phi_{2e-fl}(T)> =0$, but  $<\Phi_{2e-fl}({\bf r}) \Phi_{2e-fl}({\bf r}')> (T)$ has algebraically decaying fluctuations in $({\bf r -r'})$ below $T^{mf}_{2e, \hat{2}, \hat{3}}$ up to the transition temperature of the 2e state denoted by $T_{2e}^c$; the fluctuating region may be as large as $3~T_{2e}^c$, as can be inferred from the Monte-Carlo calculations \cite{Shiba}. Below $T^c_{\hat{6e}}$, the state $\Psi_{\hat{6}}$ co-exists with it. But in view of the fact noted above that the 
state $\Psi_{\hat{6}}$ cannot be 
absolutely stable and decays to effectively normal states, the conductances of the co-existing states are in parallel or their resistance is in series, i.e.
\be
G^{-1} = G^{-1}_{fluc-2e} + G^{-1}_{\hat{6}},
\ee
\be
\label{R}
R = R_{fluc-2e} + R_{\hat{6}}.
\ee
Therefore in a bulk or ring geometry without a field $R(T) = R_{fluc-2e}(T)$, and is finite and varying in temperature till $T \leqslant T_{2e}^c$.

Let us now consider the ring-geometry with a flux through it. On formation of a vortex in the two arms of the ring when the flux through the ring is 6e-flux quantum the state r $\Psi_{\hat{6}}$ ) responds so that $R_{\hat{6}}$ acquires a finite value. This happens periodically as the flux is increased to form larger number of vortices. Accordingly through Eq. (\ref{R}), the resistivity rises periodically over the resistivity given by $R_{fluc-2e}(T)$. Note that this is precisely what happens in the experiment - resistivity increases periodically over a temperature dependent value; it never goes to zero. This is different from the usual Little-Parks oscillations in two ways. The usual oscillations oscillate between zero and a finite value and are confined to a small temperature region near the transition temperature because the transition temperature moves periodically as vortex (with a flux quantum) is formed in  the geometry.

In the experiments \cite{JiangWang-arXiv22}, 4e oscillations with smaller amplitude occur at a temperature below where the 6e oscillations begin to be observed.
They also terminate when true long-range order occurs in the 2e oscillating state. Based on the estimate of the mean-field transition temperature, we would expect the 4e oscillations to start at a higher transition temperature in a BCS theory. But we are dealing with Kosterlitz-Thouless (KT) transitions. The KT transition temperature for a state with quantized circulation $|\kappa|$  are given by equating the characteristic energy of interaction of a pair of oppositely charged vortices of density $\rho$, which is $\kappa^2 \rho \log \rho$ with the free-energy contribution due to their entropy $T S$ which depends on their density, but not their $\kappa$,    $S(\rho) = \rho \log \rho$.
This gives
\be
T_{KT}(\kappa) \propto \kappa^2.
\ee
above which vortices of quantization $\kappa$ freely proliferate in the disordered state. The ratio of $\kappa$ for 6e states is 3/2 times that for the 4e state.  So in the simplest consideration, the Kosterlitz-Thouless transition temperature for the 6e state, $T^c_{\hat{6}e}$ is $9/4$ times larger than that for the 4e state, i.e. $T^c_{\hat{6}e}$. The actual estimate may vary in better calculations. All of this is sketched in Fig. (\ref{Fig:Temperatures}).

We should note the paper by Pan and Lee (Phys. Rev. 106, 184515 (2022)) which shows that for fluctuating or decaying superconducting states in space and time, the path though the thick ring in the experimental geometry of \cite{JiangWang-arXiv22} diffusively hugs the inner parts of the ring. This justifies the use of the area of the inner part of the ring to calculate the value of the quantization. That paper also gives the limits on the lifetime of the fluctuations in relation to the size of the ring that quantization may be observed. The time-scales are very hard to estimate; this part is not addressed there or here and so the quantitative conditions for the oscillations remain undecided.

The experiments \cite{JiangWang-arXiv22} show three rounded steps in the resistivity. This is in consonance qualitativelly with the theory here and the schematic temperatures given in Fig. (\ref{Fig:Temperatures}), with the first step at $T^{mf}_{2e, \hat{1},\hat{2}}$, where the resistivity sharply diminishes due to the superconducting fluctuations of the frustrated 2e state, the second at $T^c_{\hat{6}}$ and the third at $T^c_{\hat{4}}$, where the transitions to the 6e and the 4e states occur in the fluctuating regime of the 2e state. The 2e state itself appears to condense without a sharp resistivity drop. We suspect that the details of how the 6e and 4e states disappear as the 2e state condenses and the temperature dependence of transport and thermodynamic properties near $T^c_{2e}$ pose an interesting theoretical problem (not tackled here) as well as an interesting experimental challenge to decipher.

After this paper was finished, evidence has been presented \cite{T-break-expts2023} that the low temperature superconducting state is time-reversal breaking as well as Chiral. The time-reversal breaking is consistent with our prediction. The chirality is consistent if the normal state has loop-current order.

\section{Summary and Concluding Remarks}

The kagome superconductors have a complex multiband electronic structure with multiple Fermi surfaces.
In this work, we considered the simplest, minimal single-orbital model on the kagome lattice. The model
captures the most essential feature of the 
electronic structure: a kagome band derived from hybrid $d$-electron orbitals with its p-type van Hove singularity located close to the Fermi level \cite{ZhouWang}. Despite the simplification, the model has been shown
to produce 3Q CDW states driven by extended Coulomb interactions, including both the real CDW with an inverse Star-of-David bond configuration, and the complex CDW with loop current order that breaks time-reversal symmetry \cite{ZhouWang,tv1v2}. The theory predicts the six reconstructed Fermi surface pockets in both cases in the $2\times2$ ordered state as shown in Fig.1(c), which have been observed recently by ARPES and STM experiments \cite{ilija-pocket}.

We studied such a model of six Fermi pockets for the phase fluctuations above its charge-2e chiral superconducting state and mapped the problem to a frustrated antiferromagnetic XY model on a triangular lattice. 
This model has not been amenable to analytical calculations, as far as we know. We have used Monte-Carlo results obtained many years ago \cite{Shiba,  Sudbo2014} to argue for an extended region of fluctuations of the 2e state due to frustration. In contrast, there is no frustration for the charge 6e and 4e flux quantized states that we have introduced since they have no internal phases. The relative transition temperature of the 6e and the 4e states is estimated using the simplest idea proposed by KT for the transition temperature. Their relation to the actual 2e transition is also estimated. There can  be no transition to the 6e or 4e states, because, as show their free-energy below their KT transition temperatures is higher than the co-existing orthogonal fluctuating frustrated 2e state. The state of the system is written as the product of such orthogonal states. The resistivity in this situation is the sum of the resistivity of the co-existing orthogonal states. In the geometry of the ring, the free-energy of the 6e and 4e states oscillates and therefore their transition temperature oscillates at their characteristic quantization period as  the flux through the ring changes. Therefore over a gradually decreasing resistivity due to the onset of the 2e  superconducting fluctuations, there are oscillations with the 6e and the 4e periods. These findings agree with recent experiments probing the superconducting properties of CsV$_3$Sb$_5$ \cite{JiangWang-arXiv22, Kapitulnik-arXiv23, T-break-expts2023} for the fluctuation region, the flux quantizations and their order as the temperature decreases.
While a full account of these experimental discoveries may require taking into account the complex band structure, our findings based on the simplified model provide a plausible physical mechanism with which we hope to stimulate further experimental and theoretical investigations. 

Although a fairly complete account of the extended fluctuation regime has been given in this paper, the conditions for the occurrence of Little-Parks oscillations with $6e$ and $4e$ flux-quantization are only only qualitatively given. This is due to the fact that the oscillations are in the fluctuating regime and the quantitative details depend on lifetime of states, which are very hard to estimate.

The general considerations here should apply to any three band or three-component superconductor in which three different paired states are weakly coupled; this is really a phenomena having to do with the critical fluctuations which turn into multiple fluctuating Leggett modes in the superconducting state. It is noteworthy that the Fe-based superconductor Ba$_{1-x}$K$_x$Fe$_2$As$_2$ \cite{Babaev, Sudbo2013, Sudbo2014} which is expected from its band-structure to be a three band-superconductor shows in its resistivity and specific heat a very extended region of fluctuations just as in CsV$_3$Sb$_5$. We suggest Little-Parks experiments for it also to see if in the fluctuation regime 6e and 4e  flux-quantization can be observed.

A question which is easy to answer is what happens to two band superconductors which have only one internal phase and one Leggett mode. In that case, in equilibrium the internal phase acquires the value $0$ so that the usual Ginzburg-Landau equations are obtained. However, three or more bands/components superconductivity with weak coupling among different pairing states will in general have unusual fluctuations.

\section{Acknowledgments}
We thank Aspen Center for Physics for hospitality and acknowledge the support of NSF Grant No. PHY-1067611. ZW is supported by the U.S. Department of Energy, Basic Energy Sciences (Grant No. DE-FG02-99ER45747) and by Research Corporation for Science Advancement (Cottrell SEED Award No. 27856).

\end{document}